# Transport spectroscopy for Paschen-Back splitting of Landau levels in InAs nanowires

Bum-Kyu Kim[1][¶], Sang-Jun Choi[2][¶], Jae Cheol Shin[3][¶], Minsoo Kim[4], Ye-Hwan Ahn[1,5], H.-S. Sim[2][*], Ju-Jin Kim[6][*], and Myung-Ho Bae[1,7][*]

[1]Korea Research Institute of Standards and Science, Daejeon 34113, Republic of Korea

[2]Department of Physics, Korea Advanced Institute of Science and Technology, Daejeon 34141, Republic of Korea

[3]Department of Physics, Yeungnam University, Gyeongsan 38541, Republic of Korea

[4]Department of Physics, Pohang University of Science and Technology, Pohang 790-784, Republic of Korea

[5]Department of Physics, Korea University, Seoul 136-713, Republic of Korea

[6]Department of Physics, Chonbuk National University, Jeonju 561-756, Republic of Korea

[7]Department of Nano Science, University of Science and Technology, Daejeon, 34113, Republic of Korea

[¶]These authors contributed equally to this work.

[*]Corresponding authors: hssim@kaist.ac.kr, jujinkim@chonbuk.ac.kr, mhbae@kriss.re.kr

## Abstract

The coupling of electron orbital motion and spin leads to nontrivial changes in energy-level structures, leading to various spectroscopies and applications. In atoms, such spin-orbit coupling (SOC) causes anomalous Zeeman splitting, known as the Paschen-Back (PB) effect, in the presence of a strong magnetic field. In solids, SOC generates energy-band inversion or splitting, a prerequisite for topological phases or Majorana fermions, at zero or weak magnetic fields. Here, we present the first observation of PB splitting of Landau levels (LLs) in indium arsenide nanowires in a strong-field regime. Our energy-resolved transport spectroscopy results indicated the presence of LL-dependent anomalous Zeeman splitting in these nanowires, analogous to the atomic PB effect. This result was found to be in good agreement with a theoretical analysis based on Rashba SOC. Our findings also suggested a way of generating spin-resolved electron transport in nanowires.

Landau orbitals, circular motions of an electron in a two-dimensional electron gas (2DEG) system under a magnetic field, give rise to spectra showing discrete energy levels called Landau levels (LLs). The energy levels are further split when the degrees of freedom of spin, valley, or the number of layers, or electron-electron interactions, come into play.[1,2,3,4,5] LL spectroscopy has provided valuable information about this splitting and suggested the possibility of new phases of matter. The interplay between spin-orbit coupling (SOC) and Zeeman effects can also be expected to generate such splitting. Indeed, a beating pattern in Shubnikov-de Haas oscillations in an InGaAs/InAlAs 2DEG system[6] and a nonlinear spin splitting of LLs in an InSb 2DEG system[7] under a relevant magnetic field have been attributed to this type of interplay.

In our work, we study the interplay in the regime of an effectively stronger magnetic field where the SOC strength is regarded as a perturbation to the Zeeman energy, in an analogy to the atomic Paschen-Back (PB) effect.[8,9,10] To identify the interplay in the strong magnetic field limit, we use indium arsenide (InAs) nanowires, which show strong Rashba SOC (RSOC),[11] and apply a gate voltage to the nanowires. In this case, the RSOC strength in InAs nanowires is one order of magnitude stronger than that of typical semiconducting heterostructures;[12,13,6] moreover, charge depletion by the gate voltage can give rise to the formation of a 2DEG in the surface region of the nanowires.[14] Note that most previous works[15,14,16,17,18] on nanowires reported electron transport behavior at a zero or weak magnetic field, which was explained in these reports by one-dimensional subbands, rather than LLs. A study of such nanowires under a strong magnetic field has been reported[13], but that report did not include an analysis of the effects of LLs. In the current work, we theoretically showed the Zeeman splitting of LLs under a strong magnetic field to change due to RSOC in an anomalous fashion, depending on the quantum numbers of the LLs and the electron spin. We refer to this resulting splitting as the PB splitting of LLs. We experimentally confirmed the anomalous Zeeman splitting of LLs by performing energy-resolved transport spectroscopy on the InAs nanowires.

**RESULTS**

First, we discuss the PB splitting of LLs. Landau orbitals of an electron in a 2DEG under a perpendicular magnetic field $B$ are affected by the Zeeman energy and the RSOC (see Figure 1(a)). The Hamiltonian for this system can be written[11] as $H = \left(\hat{a}^\dagger\hat{a} + \frac{1}{2}\right)\hbar\omega_c + \frac{1}{2}\Delta_Z\sigma_z - \frac{\alpha}{\sqrt{2}l_B}(\hat{a}\sigma_+ + \hat{a}^\dagger\sigma_-)$. Here, $\omega_c = eB/m^*$ is the cyclotron frequency, $m^*$ the effective mass of the electron, $e$ the electron charge, $\Delta_Z = g\mu_B B$ the Zeeman energy, $g$ the Landé g-factor, $\mu_B$ the Bohr magneton, $\sigma_{x,y,z}$ the Pauli matrices describing the electron spin, $\sigma_\pm = \sigma_x \pm i\sigma_y$, $\alpha$ the RSOC strength, $l_B = \sqrt{\hbar/eB}$ the magnetic length, and $\hbar = h/2\pi$ the Planck constant. The operator $\hat{a}\sigma_+$ ($\hat{a}^\dagger\sigma_-$) lowers (raises) the LL orbital quantum number $n$ ($= 0,1,2,...$) by 1 and raises (lowers) the electron spin-z angular momentum number $\sigma/2$ ($= 1/2$ for the spin-z state ↑ and -1/2 for ↓) by 1. When the magnetic field is sufficiently strong ($B > 3$ T for InAs nanowires), the RSOC can be considered as a perturbation in comparison with $\hbar\omega_c$ and $\Delta_Z$. In this regime, we determined the energy of electrons with quantum numbers $n$ ($= 0,1,2,\dots$) and $\sigma$ according to eq 1 (see Section 1 of Supporting Information)

$$E_{(n,\sigma)} \approx \left(n + \frac{1}{2}\right)\hbar\omega_c + \sigma\frac{\Delta_Z}{2} - \sigma\Delta_{PB}(n,\sigma), \qquad \Delta_{PB}(n,\sigma) \equiv \frac{2(2n+\sigma+1)}{1 - \frac{\Delta_Z}{\hbar\omega_c}}\Delta_{so}, \qquad (1),$$

where $\Delta_{so} \equiv \frac{m^* a^2}{2\hbar^2}$. Eq (1) shows the dependence of the anomalous Zeeman splitting of the LLs on $n$ and $\sigma$. This corresponds to the atomic PB effect, where the electrons in an atom exhibit Zeeman splitting that depends on their orbital and spin angular momentum quantum numbers due to their SOC.

Figure 1(b) shows the energy levels $E_{(n,\sigma)}$ of Eq. (1) in three regimes, specifically (i) $\Delta_Z = \Delta_{so} = 0$, (ii) $\Delta_Z \neq 0$, $\Delta_{so} = 0$, and (iii) $\Delta_Z \gg \Delta_{so}$. In Regime (ii), the Zeeman effect lifts the spin degeneracy of LLs by $\Delta_Z$. In PB Regime (iii), the dressed RSOC energy $\Delta_{PB}$ results in an additional energy shift, as follows: $\Delta_{PB}(n,\sigma)$ satisfies $\Delta_{PB}(n+1,\sigma=\downarrow) = \Delta_{PB}(n,\sigma=\uparrow)$, as $n + \frac{\sigma}{2}$ represents the constant motion of the Hamiltonian $H$. Hence, the RSOC causes level repulsion between the $(n+1, \sigma=\downarrow)$ and $(n, \sigma=\uparrow)$ states. As a result, the Zeeman splitting in each LL with $n \geq 1$, $\Delta E_n \equiv \Delta_Z - \Delta_{PB}(n,\uparrow) - \Delta_{PB}(n,\downarrow)$ is reduced from the bare Zeeman splitting $\Delta_Z$, while it is $\Delta E_0 \equiv \Delta_Z - \Delta_{PB}(0,\uparrow)$ for the lowest ($n = 0$) LL given that $\Delta_{PB}(n=0,\sigma=\downarrow) = 0$. Also note that $\Delta E_0 > \Delta E_1$.

Figures 1(c) and 1(d) show the electron density of states (DOS) and the resulting two-terminal electron conductance $G$ of the 2DEG in Regimes (ii) and (iii), respectively. The Fermi energy $E_F$ is tuned by the gate voltage ($V_g$) applied to the 2DEG. Each spin-resolved LL gives rise to a jump by $G_0/2 \equiv e^2/h$ in the conductance as a function of $V_g$ whenever the Fermi energy crosses the LL[19] (see the jumps along the vertical dashed line in Figure 1(c)). A conductance jump also occurs as a function of the source-drain bias voltage $V_{sd}$ whenever $V_{sd}$ increases such that the LL starts to contribute to the conductance. As a result, diamond patterns appear in the conductance map, drawn as a function of $V_g$ and $V_{sd}$. As shown in the right panels in Figures 1(c) and 1(d), the conductance map of the PB Regime (iii) is distinct from that in Regime (ii). For instance, the size of $\Delta_Z$ of the green diamond is identical to that of the purple diamond in Regime (ii), while these two diamonds have different sizes, corresponding to $\Delta E_0$ and $\Delta E_1$, in PB Regime (iii). Therefore, transport spectroscopy is a good tool for identifying the PB splitting of LLs in experiments.

To observe the PB splitting of LLs, we measured the electrical conductance through InAs nanowires. Here, the InAs nanowires were grown using the catalyst-free metalorganic chemical vapor deposition (MOCVD) method. The grown nanowires were deposited on 300-nm-thick $SiO_2$/Si substrates to fabricate electrical devices (see Methods and Section 2 of Supporting Information). Based on the traditional nanofabrication process, we prepared InAs field-effect transistors (FETs), as shown in Figure 2(a), where the upper and lower panels display SEM images of the devices, named by NW1 (channel length, $L \sim 85$ nm and diameter, $D \sim 70$ nm) and NW2 ($L \sim 160$ nm, $D \sim 85$ nm), respectively.

Charge depletion by the back-gate voltage $V_g$ gives rise to the formation of a quasi-2DEG at the top of the channel, located above the gate (see Figure 2(b)).[14] This was supported by a numerical simulation (see Section 3 of Supporting Information). The width of the 2DEG in this case was expected to be comparable to the width, ~45 nm, of the side of the hexagonal cross-section of the nanowires. The 2DEG was sufficiently wide to host LLs under a strong magnetic field of $B > 4$ T, because of (1) $l_B$ (= 13 nm at $B = 4$ T, 8.7 nm at 9 T) being shorter than the width of the 2DEG and (2) momentum kicks by the Lorentz force resulting in sufficient mixing of the one-dimensional subbands of the nanowires (see Section 1 of Supporting Information).

As a prerequisite for the PB splitting of LLs, we experimentally confirmed the formation of LLs in the InAs nanowires. Figure 3(a) shows the zero-bias conductance of NW1 as a function of the gate voltage ($G$-$V_g$ curves). Here, $B$-fields ranging from zero to 9 T were applied perpendicular to the substrate (see Figure 2(b)) at $T = 50$ K. By increasing the temperature up to 50 K, we suppressed the Fabry-Perot interference and hence clearly observed the LL-induced conductance steps (see Section 4 of Supporting Information).[20,21] At $B = 4$ T, a new conductance step emerged near $G \sim 0.7$, as indicated by the vertical arrow in Figure 3(a), and this step became wider without its $G$ value changing as we increased the $B$-field to $B = 9$ T, as indicated by the horizontal arrow. To reveal the origin of the observed conductance step, we produced a map in which we plotted d$G$/d$V_g$ as a function of both $V_{sd}$ and $V_g$ under $B = 5$, 7 and 9 T (from left to right as shown in Figure 3(b)). The structure of the conductance step in the $G$-$V_g$ curves (Figure 3(a)) corresponded to the diamond-shaped structure in each map (dashed lines of Figure 3(b)). $\Delta V_{sd}$, defined as half of the width of the diamond-shaped structure, was observed to increase as the $B$-field was increased (Figure 3(b)). The squares in Figure 3(c) show $\Delta V_{sd}$ as a function of the $B$-field for NW1. For $B > 5$ T, the experimentally obtained energy scales of $\Delta V_{sd}$ were found to be in good agreement with $E_{(1,\downarrow)} - E_{(0,\downarrow)}$ (solid red line) as obtained from eq (1) with parameters[12,22,23,24,25] $\alpha \sim 0.36$ eV·Å, $g = 20$ and $m^*$=0.025$m_e$, where $m_e$ is the rest mass of a free electron. This agreement indicated that the conductance steps observed in the $G$-$V_g$ curves for $B > 5$ T in Figure 3(a) originated from the LLs. The conductance value corresponding to the conductance step was ~0.7, not 1, and this value was attributed to the location of 2DEG partly in the middle of the channel, with inevitable contact resistance. For NW1, we could not observe the Zeeman splitting of the LLs, due to the high temperature of $T = 50$ K with respect to the Zeeman-splitting energy scale ($\Delta_Z \sim 10.4$ meV at $B = 9$ T).

At this point, to elucidate the PB splitting of the LLs, we carried out transport spectroscopy at a lower temperature with using a longer nanowire. Figure 4(a) shows the zero-bias conductance $G$-$V_g$ curves of the relatively long InAs nanowire NW2 with various $B$-fields at $T = 35$ K. We found a robust conductance step at $G \sim 0.5$ in each of these curves, as indicated by the arrow in Figure 4(a). Moreover, in the d$G$/d$V_g$ maps derived from the data for NW2 in Figure 4(b), the diamond-shaped structures were observed to increase in size as the $B$-field was increased. This behavior was also indicated by the data corresponding to the circles in Figure 3(c) and also found to be consistent with $E_{(1,\downarrow)} - E_{(0,\downarrow)}$ for $B > 7$ T.

As marked by the dashed red arrows in the right panel of Figure 4(b), a sub-structure was observed in the diamond-shaped structure of the d$G$/d$V_g$ map derived from the B = 9 T data for NW2. We suggest that the sub-structure corresponded to the energy splitting of the lowest LL. The energy splitting $\Delta E_0$ was deduced from the d$G$/d$V_g$ map (Figure 4(b)) to be $\sim 12 \pm 1$ meV, comparable to the predicted $\Delta E_0$ value of $\sim 9.4$ meV based on eq (1) considering a thermal broadening $k_B T$ of $\sim 3$ meV at $T = 35$ K. In contrast, the energy splitting of the first LL ($n$=1) $\Delta E_1$ was obscured, as indicated by two yellow dashed lines for both $V_{sd}$ polarities. This result was due to a thermal broadening smearing out the energy splitting of $\Delta E_1 \equiv \Delta_Z - \Delta_{PB}(1,\uparrow) - \Delta_{PB}(1,\downarrow)$ of $\sim 7.2$ meV. To verify the observed sub-structure, we performed a numerical simulation of the d$G$/d$V_g$ map, considering the LL, Zeeman and RSOC effects in the InAs 2DEG system at $T = 35$ K with a series resistance $R_s$ of 10 k$\Omega$. The left and right panels of Figure 4(c) show the simulated plots at $B = 4$ T and 9 T, respectively, considering the dressed RSOC in addition to the LL and Zeeman interaction, i.e., in PB Regime (iii). The left and right panels of Figure 4(d) show the results of the theoretical simulations at $B = 4$ T and 9 T, respectively, with only the LLs and Zeeman interaction, i.e., in Regime (ii) (see Figure 1(b)). Although the results at $B = 4$ T were similar in both cases, the plots at $B = 9$ T without and with the dressed RSOC clearly differed (see also Figures 1(c) and 1(d), correspondingly). In Regime (ii) (the right panel of Figure 4(d)), identical amounts of energy splitting by $\Delta_Z$ were calculated for the LLs with $n = 0$ and $n = 1$. However, in PB Regime (iii) in the right panel of Figure 4(c), the dressed RSOC significantly suppressed the Zeeman splitting at the $n = 1$ LL, which was almost smeared out at $T = 35$ K. This behavior is consistent with the experimental observation shown in the right panel of Figure 4(b).

To analyze these results in detail, sets of $dG/dV_g$-$V_g$ curves, shown in Figures 5(a)-(c), were obtained by slicing the experimentally derived and two numerically calculated $B = 9$ T $dG/dV_g$ maps (shown in the right panels of Figures 4(b)-(d), respectively) at various $V_{sd}$ values. The curves are vertically shifted from one another in the figure for clarity, and the red curves represent the data obtained at zero bias. The three sets of $dG/dV_g$-$V_g$ curves showed similar patterns of dips (crossing dashed lines) originating from the $n = 0$ LL. Such behavior was also found at the $n = 1$ LL region for the set of $dG/dV_g$-$V_g$ curves (Figure 5(c)) derived from the map calculated without the dressed RSOC, i.e., in Regime (ii), but not for the set of $dG/dV_g$-$V_g$ curves (Figure 5(b)) corresponding to PB Regime (iii), and not for the set (Figure 5(a)) derived from the experimentally obtained map. These results indicated the experimental $B = 9$ T $dG/dV_g$ map to be consistent with the PB Regime (iii). This consistency was also evident when comparing the peak amplitudes of non-shifted $dG/dV_g$-$V_g$ curves for positive $V_{sd}$ values only (Figure 5(d)). In the $n = 1$ LL region, the curves (right panel of Figure 5(d)) derived from the map calculated without the dressed RSOC, i.e., in Regime (ii) showed two extra peaks, as indicated by the arrows, but these peaks were strongly suppressed in the two other sets of curves (panels on the left and in the middle in Figure 5(d)). The same behavior was also observed for the negative $V_{sd}$ cases (see Section 5 in Supporting Information). Thus, we conclude the experimental results to be consistent with the theoretical result considering the effect of the dressed RSOC, rather than that with only the Zeeman interaction.

**CONCLUSIONS**

In summary, our energy-resolved transport spectroscopy elucidated the PB splitting of LLs in a 2DEG formed in InAs nanowires by a relevant back-gate voltage condition. This spectroscopy is expected to be applicable to other nanowires with strong SOC, such as a $Bi_2Se_3$ topological insulator.[26,27,28,29] Our findings represent progress toward achieving spin-resolved quantum Hall effects in nanowires.

**Methods**

**InAs nanowire growth** The InAs nanowires used in this study were grown in a MOCVD system with a horizontal reactor (A200/AIXTRON Inc.). A two-inch Si(111) wafer was cleaned with buffered oxide etchant for 1 min and rinsed in deionized (DI) water for 5 sec. Then, the wafer was dipped in a poly-L-lysine solution (Sigma-Aldrich Inc.) for 2 min and rinsed in DI water for 10 sec. The wafer was then loaded into the MOCVD reactor. To grow the nanowires, the reactor pressure was lowered to 50 mbar with 15 L/min of hydrogen gas flow and the reactor temperature was increased to 570 °C. After a short stabilization time, arsine gas ($AsH_3$) was flowed into the reactor for 1 min, followed by trimethylindium (TMIn). The molar flow rates (mol/min) of TMIn and $AsH_3$ were $2\times10^{-5}$ and $2\times10^{-4}$, respectively. The nanowire growth was performed for 90 min to form 20-μm-long InAs nanowires. The morphologies of the InAs nanowires were investigated using scanning electron microscopy (SEM, Hitachi-S4700, Hitachi Inc.). The structural properties of the InAs nanowires were examined using transmission electron microscopy (JEM-2200FS, JEOL Inc).

**Sample fabrication** To fabricate the InAs nanowire FETs, as-grown InAs nanowires were dispersed in isopropanol by carrying out sonication for 10 sec, and then dispersed on a highly doped silicon substrate covered by a 300-nm-thick oxide layer with pre-patterned metal markers. The highly doped silicon substrate was used as the global back-gate electrode. The multi-terminal electrodes were defined using standard electron-beam lithography and electron-beam evaporation of Ti/Au (10/150 nm). Prior to metal deposition, the native oxide layer on the InAs nanowire surface was etched in a 2% diluted solution of 3 M ammonium polysulfide $(NH_4)_2S_x$ at 60 °C for 5 min to achieve low contact resistance. For the electrical characterizations, we measured the differential conductance using a standard lock-in amplifier (SR830) with an excitation voltage of 100 μV and a frequency of 1.777 kHz superimposed on a DC bias voltage (Yokogawa 7651) in the physical property measurement system (PPMS, Quantum Design, Inc.).

## ACKNOWLEDGMENTS

This work was supported by the Korea Research Institute of Standards and Science (KRISS 2017 GP2017-0034), part of the Basic Science Research Program through the National Research Foundation of Korea (NRF) (Grant Nos. 2015R1A2A1A10056103, SRC2016R1A5A1008184, 2016R1A2B4008525 and 2017R1C1B2010906). This work has also been partly supported by the Korea-Hungary Joint Laboratory for Nanosciences programme through the National Research Council of Science and Technology.

**Figure captions**

**Figure 1.** (a) Landau orbit of an electron in a 2DEG, defined to be on an *xy* plane, under a perpendicular magnetic field $\boldsymbol{B} = B\hat{\boldsymbol{z}}$. In this set-up, the Landau orbital is affected by the Zeeman effect of the Hamiltonian $H_{ZM} \propto \boldsymbol{B} \cdot \boldsymbol{\sigma}$ and the RSOC of $H_{SO} \propto \boldsymbol{B}_{SO} \cdot \boldsymbol{\sigma}$, where $\boldsymbol{B}_{SO} \propto \boldsymbol{v} \times \boldsymbol{E}$ is the effective magnetic field induced by the RSOC, $\boldsymbol{\sigma}$ describes the electron spin, $\boldsymbol{v}$ the electron velocity, and $\boldsymbol{E}$ the electric field induced by a back-gate voltage. When $H_{ZM} \gg H_{SO}$, this situation corresponds to the atomic Paschen-Back effect where the energies of the atomic orbitals exhibit anomalous Zeeman splitting. (b) Splitting of the LLs of the 2DEG in three regimes, specifically (i) bare LLs, (ii) LL+ZM (LL splitting by Zeeman energy $\Delta_Z$) and (iii) LL+ZM+SOC (LL splitting by the Zeeman effect and the RSOC in the PB regime of a strong $B$). Electron state are characterized by the LL orbital quantum number $n$ and spin z quantum number $\sigma$. $\Delta E_0$, $\Delta E_1$, and $\Delta_{PB}(n, \sigma)$ are the energy splitting of the $n = 0$ LL, that of $n = 1$, and the dressed RSOC energy in PB Regime (iii). (c) Schematic DOS (left panel) and conductance map (right) of the two lowest LLs ($n = 0, 1$) of a 2DEG in Regime (ii). The DOS shows spin splitting by $\Delta_Z$. The map shows the two-terminal conductance G of the 2DEG as a function of source-drain voltage $V_{sd}$ and gate voltage $V_g$. (d) The same as (c), but in PB regime (iii). The spin splitting energies $\Delta_Z$, $\Delta E_0$, and $\Delta E_1$, were identified from the size of the diamond structures in the map.

**Figure 2.** (a) Upper and lower panels: false-colored SEM images of NW1 and NW2, respectively. Scale bars: 1 μm. Yellow regions indicate metal electrodes and the InAs channels located between them. (b) Schematic of charge depletion in an InAs-nanowire channel by the electric field induced by a back-gate voltage. The green frame depicts the outline of the nanowire. The part of the schematic shown in grey represents the charge depletion region, and the red box indicates the region where a quasi-2DEG formed. A perpendicular magnetic field $B$ was applied to the nanowire. The direction of the electric field is indicated by the blue arrow.

**Figure 3.** (a) $G$-$V_g$ curves for the ~85 nm-long device (NW1) at various $B$-fields perpendicular to the substrate from zero (leftmost) to 9 T (rightmost) at intervals of 0.5 T, and acquired at $T = 50$ K. Data are horizontally shifted for clarity. Dashed arrows indicate step-like features in the curves, i.e., "conductance steps", observed at $B = 0$ T and that become smeared out with increasing $B$-field. These structures were attributed to Fabry-Perot interference.[30] The vertical arrow at $B = 4$ T indicates a newly emerging conductance step near $G \sim 0.7$, which became an apparent step at $B = 9$ T, as indicated by the horizontal arrow. This appearance of such a step was interpreted as a signature of the formation of the LLs in the nanowire. (b) d$G$/d$V_g$ as a function of $V_{sd}$ and $V_g$ at $B = 5$, 7 and 9 T (from left to right). The $V_{sd}$ spacing between the two vertical arrows ($2\Delta V_{sd}$) at $B = 5$ T corresponds to $2\Delta E_{10\downarrow}$. (c) $\Delta V_{sd}$ as a function of the $B$-field, where the data shown as squares and circles were obtained from NW1 and NW2, respectively. The error bar at each field was obtained by considering $\Delta V_{sd}$ values for both polarities of $V_{sd}$. The energy difference $E_{n\downarrow} = E_{n\downarrow} - E_{0\downarrow}$, obtained from eq (1), is drawn as dashed and solid lines for Regimes (ii) and (iii), respectively. The experimental $\Delta V_{sd}$ data were observed to be in agreement with $E_{n\downarrow}$

for $B > 5$ T; in our experimental set up, $l_B$ was designed to be comparable to the nanowire width in order to form well-developed LLs.

**Figure 4.** (a) $G$-$V_g$ curves for the ~165-nm-long device (NW2) at various $B$-fields and acquired at $T = 35$ K. Data are horizontally shifted for clarity. A robust step in the curves at $G \sim 0.5$, as indicated by the arrow, was observed for $B = 3$ T-9 T. (b) Experimentally determined $dG/dV_g$ maps as a function of $V_{sd}$ and $V_g$ at $B = 4$ T (left panel) and 9 T (right panel). The diamond-shaped sets of dashed lines correspond to the robust conductance steps in the $G$-$V_g$ curves shown in panel (a). In the plot at $B = 9$ T, another distinct sub-structure was observed in the vicinity of the location marked by the diamond-shaped structure, as indicated by two dashed red arrows. Black arrows indicate single $dG/dV_g$ lines at the $n = 1$ LL region. $\Delta E_0$ is shown in the right panel. (c) Numerically calculated $dG/dV_g$ maps as a function of $V_{sd}$ and $V_g$ at $T = 35$ K, considering the LL and ZM splitting with SOC (PB Regime (iii)) in an InAs 2DEG system under $B = 4$ T (left panel) and 9 T (right panel) at $T = 35$ K. For these calculations, we used $W = 44$ nm, $\alpha \sim 0.36$ eV·Å, $g = 20$ and $m^* = 0.025 m_e$. The diamond-shaped structure marked by the dashed lines corresponds to the $n = 0$ LL. Note that LL-splitting ($\Delta E_1$) at $n = 1$ was much smaller than that ($\Delta E_0$) at the $n = 0$ level. (d) Numerically calculated $dG/dV_g$ maps as in panel (c) but without the dressed RSOC, i.e., in Regime (ii). For the $B = 9$ T map, $\Delta_Z$ is denoted. Note that the level splitting due to the Zeeman effect was a constant regardless of the LL quantum number $n$.

**Figure 5.** (a)-(c) Sets of $dG/dV_g$-$V_g$ curves obtained by slicing the experimentally derived and two numerically calculated $B = 9$ T $dG/dV_g$ maps (right panels of Figures 4(b)-(d)) at various $V_{sd}$

values, specifically every 2 mV from 20 mV to -20 mV, displayed from top to bottom. The curves are vertically shifted from one another for clarity. Here, the red curves represent the results obtained at zero bias. Dips and peaks at zero bias, depicted by the polygons, correspond to features of the d$G$/d$V_g$ maps (right panels of Figures 4(b)-4(d)) marked by the corresponding polygons. Red and blue dashed curves were drawn to follow two local d$G$/d$V_g$ dips at various $V_{sd}$ values in the $n = 0$ LL region, and intersected at $V_{sd} = 0$ mV in each of the three sets of d$G$/d$V_g$-$V_g$ plots. The two smeared peaks (indicated by the triangle) in the $n = 1$ LL region of the zero-bias d$G$/d$V_g$-$V_g$ curve (panel (b)) derived from the numerically calculated map corresponding to PB Regime (iii) were also found in the corresponding curve (in panel (a)) derived from the experimental map, whereas the curves connecting local dips in the $n = 1$ LL region of the d$G$/d$V_g$-$V_g$ curves (panel (c)) derived from the numerically calculated map corresponding to Regime (ii) were found to intersect at the zero bias condition. (d) From left to right: The plots of the positive $V_{sd}$ values of the sets of d$G$/d$V_g$-$V_g$ curves in panels (a)-(c), respectively, but without the vertical shift in order to more easily compare the d$G$/d$V_g$ amplitudes in the $n = 0$ and 1 LL regions. The left and middle panels show the same trend in the modulations of the dips and peaks as those for various $V_{sd}$ values in the $n = 1$ LL region, in contrast to those in the right panel.

**Figures**

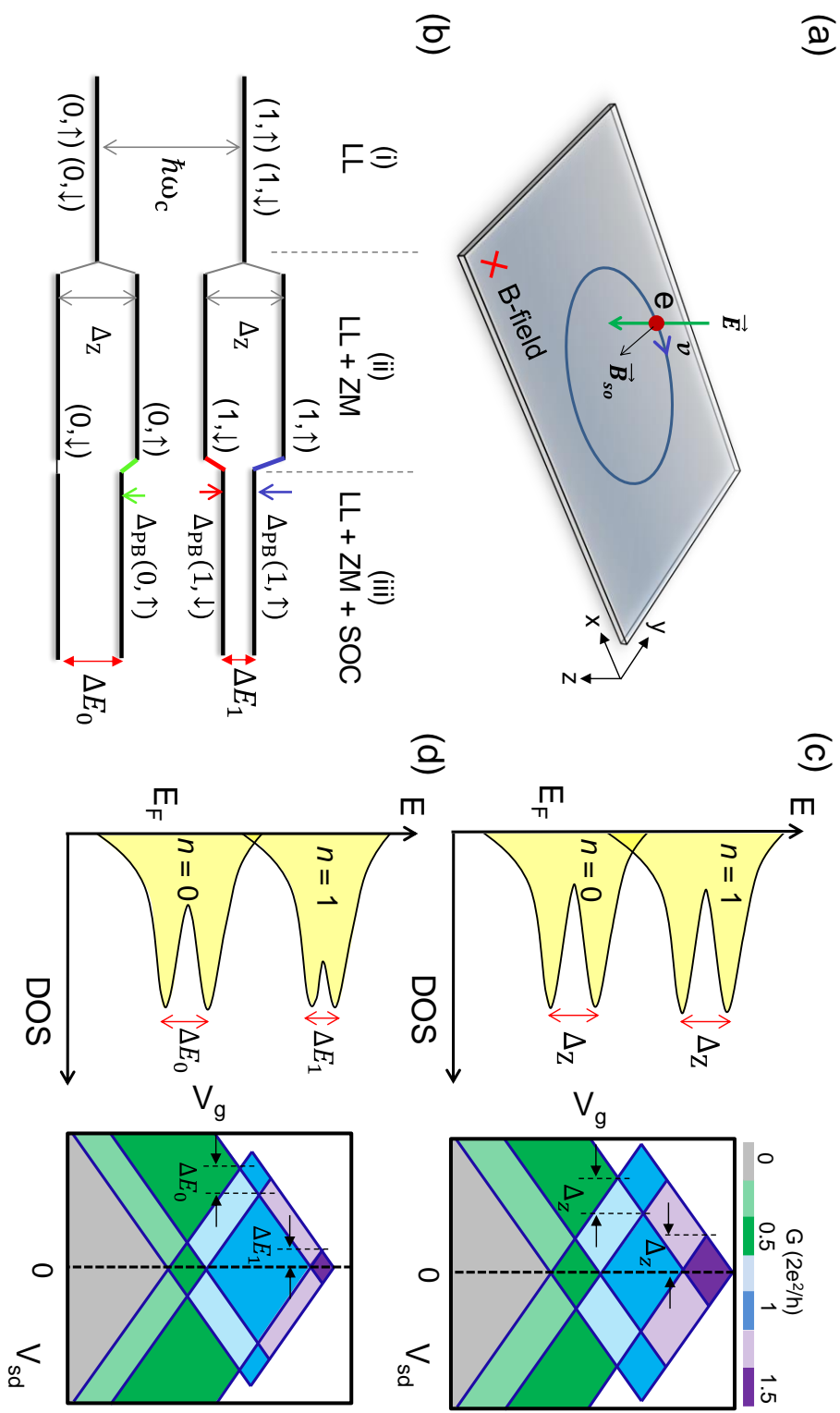

Figure 1

(a)

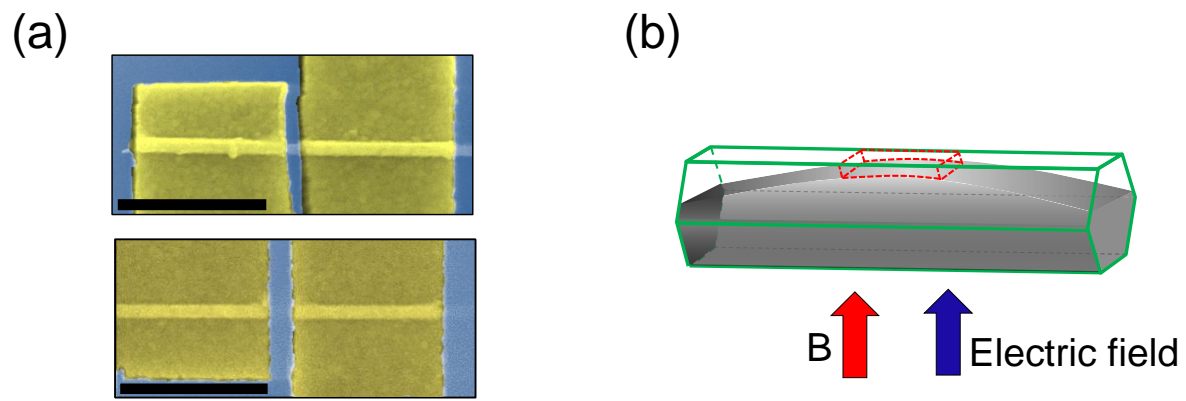

(b)

B    Electric field

Figure 2

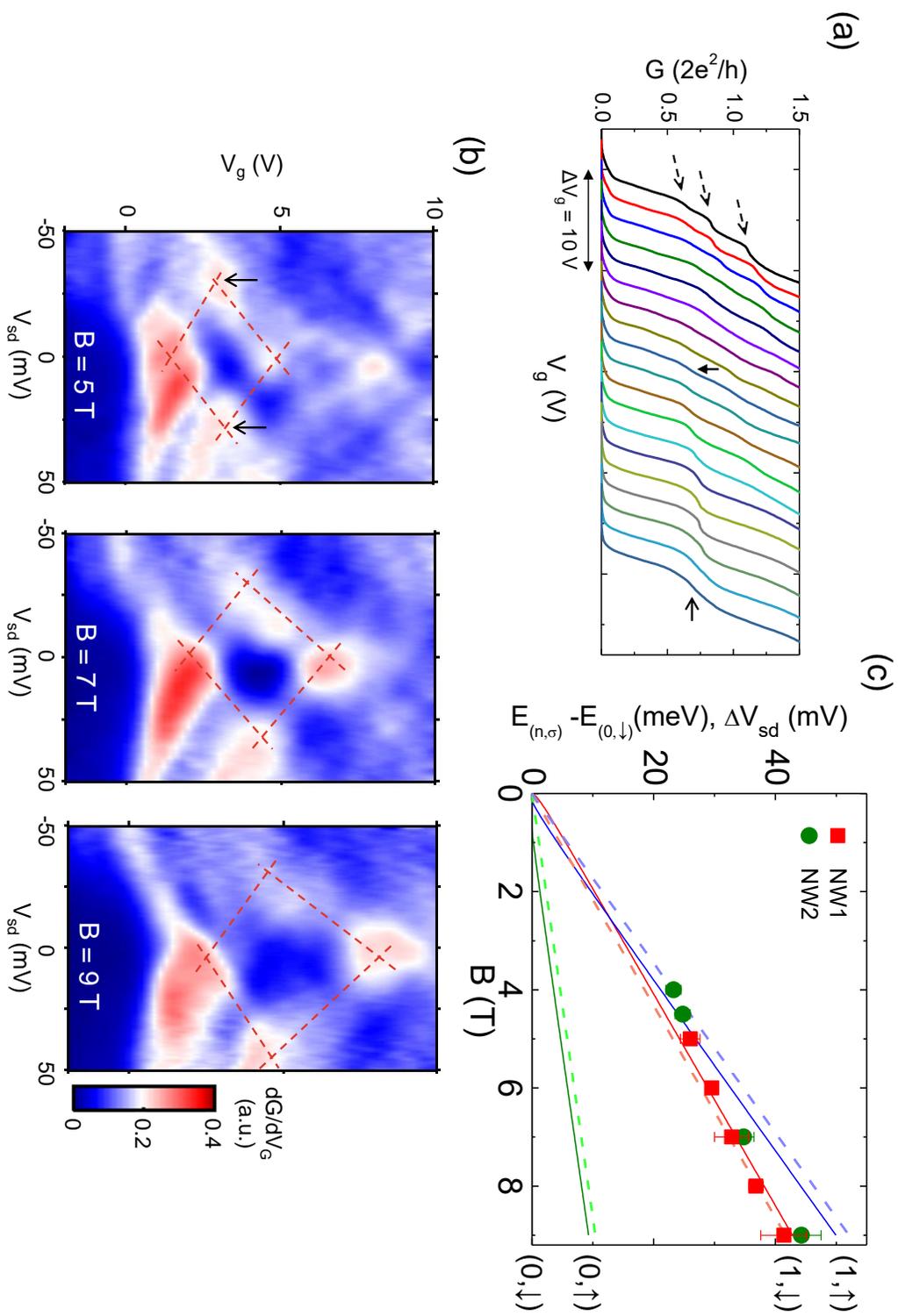

Figure 3

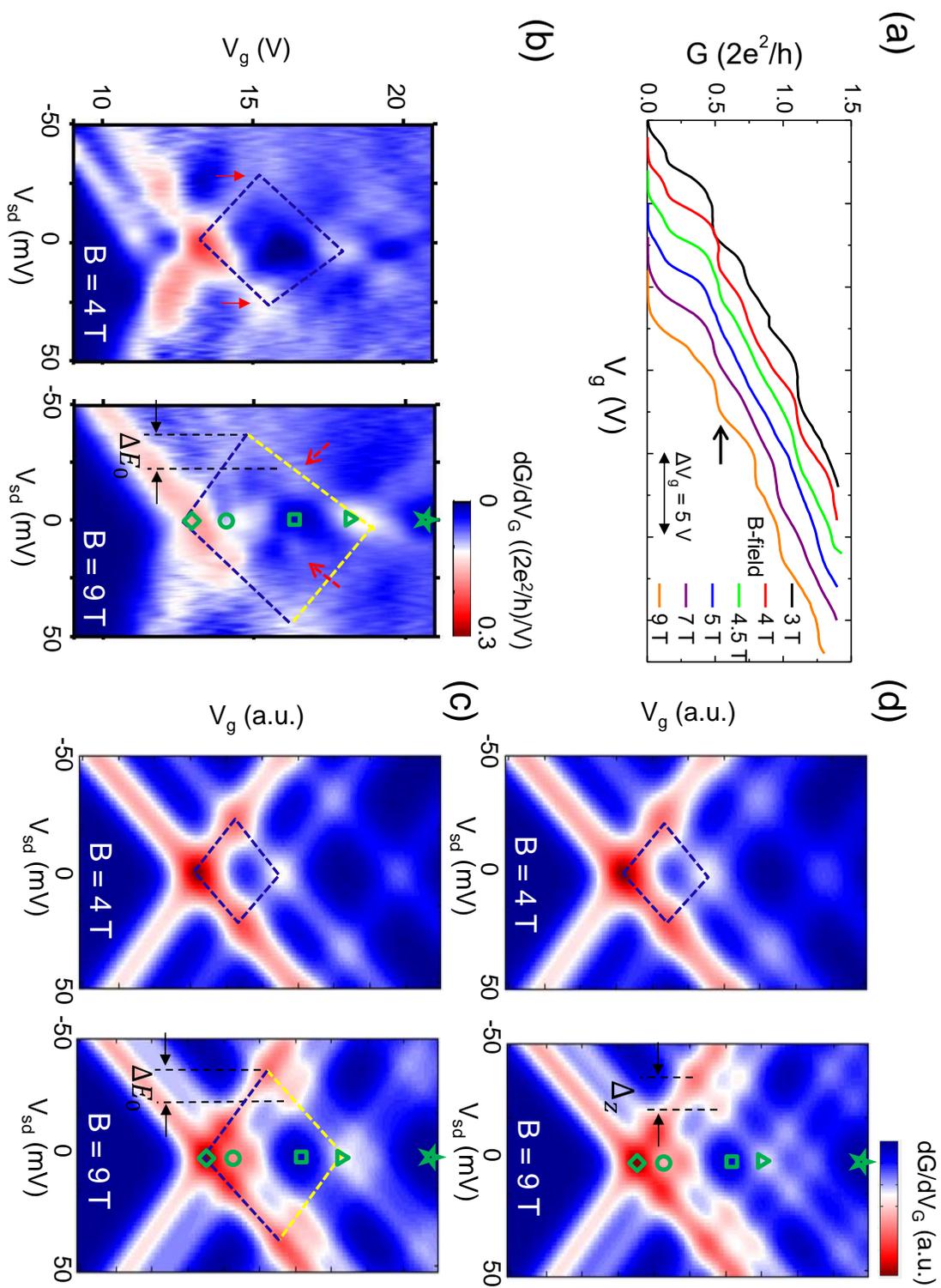

Figure 4

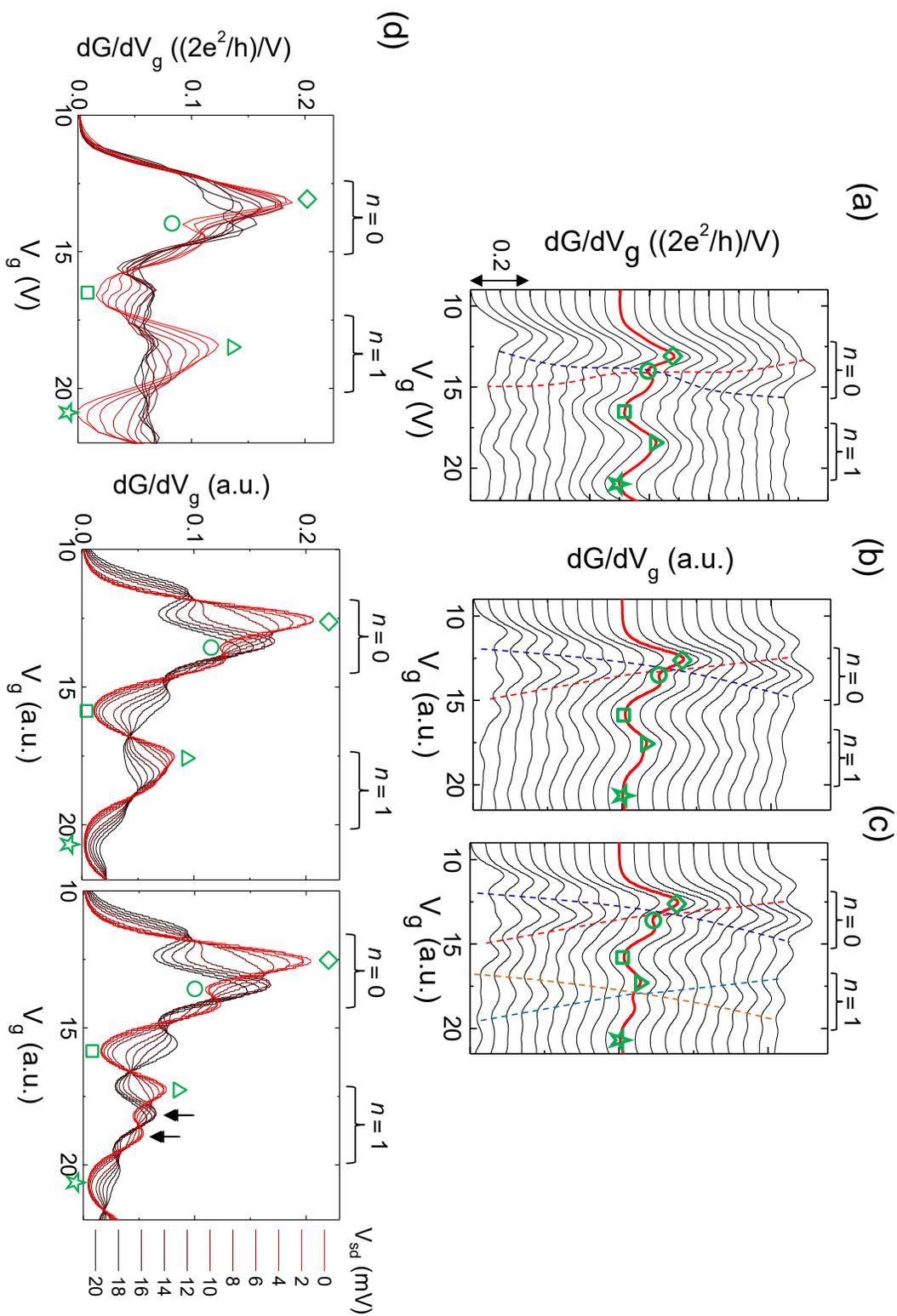

Figure 5

<div align="center">**Supporting Information**</div>

# Transport spectroscopy for Paschen-Back splitting of Landau levels in InAs nanowires

Bum-Kyu Kim, Sang-Jun Choi, Jae Cheol Shin, Minsoo Kim, Ye-Hwan Ahn, H.-S. Sim, Ju-Jin Kim, and Myung-Ho Bae

This material is composed of theory, Comsol simulation and experiment sections.

## Section 1. Theory

In this section, we first compare various scales to validate that the depleted region of our nanowires by the back gate voltage is in the two-dimensional (2D) Landau level (LL) regime at $B > 4$ T. We next present derivations of Eq. (1) in the main text. Finally, we provide a tight-binding calculation to study the interplay of the Landau level and effects of the finite lateral width $W$.

### 1. Scales and regimes

Our nanowire has the diameter of 88 nm, which is larger than that (50 nm) of conventionally studied InAs nanowires [1,2]. The following two conditions guarantee that when $B > 4$ T, 2D LLs can be formed in the depleted region of our nanowires. (i) The magnetic length $l_B$ is shorter than the lateral width $W \sim 44$ nm; at $B = 4$ T, $l_B = 13$ nm. (ii) Time scale $\Delta\tau$ for momentum kick between different one-dimensional (1D) subbands of the nanowire is much shorter than the period $\tau_c = 2\pi/\omega_c$ of the cyclotron motion so that electron motions show 2D cyclotron motion rather than 1D motion along the nanowire direction. Here, $\omega_c$ is the cyclotron frequency. In our nanowire, there are about 10 subbands at the Fermi level $E_F$, each having a quantized lateral wave vector $k_n = n\pi/W$, $n = 1, 2, \dots$ The time for momentum kick is estimated by $\Delta\tau =$ $\hbar\Delta k_n/F = \pi/(\omega_c k_F W)$, where $F = evB = \hbar\omega_c k_F$ is the Lorentz force and $k_F$ is the Fermi wave vector. Using $E_F \sim 20$ meV [3] and $m^* \sim 0.025m_e$, we find $\Delta\tau/\tau \sim 0.1$ at $B = 4$ T.

The above estimations indicate that the depleted region of our nanowire is in the regime of 2D LLs when $B > 4$ T. Indeed, the experimentally observed level spacing of $\sim 25$ meV at 4 T is certainly different from the Zeeman splitting $\sim 4.8$ meV at 4 T (estimated with Landé g-factor $\sim 20$) and the level spacing $\sim 7.8$ meV between different subbands by lateral confinement with $W \sim 44$ nm. Instead, the observed level spacing $\sim 25$ meV is comparable with the cyclotron energy $\hbar\omega_c$ at 4 T. This strongly supports the formation of 2D LLs.

Note that previous studies [4,5] on nanowires considered a different regime from ours. In those studies, a weak magnetic field is applied so that the nanowires are in the regime of quasi 1D electron transport with a few subbands. The two conditions of $l_B < W$ and $\Delta\tau/\tau \ll 1$ are not satisfied in those studies.

## 2. Derivation of Eq. (1) of the main text

Electrons under the Zeeman and Rashba SOC interactions are described by the Hamiltonian

$$H = \frac{\vec{\Pi}^2}{2m^*} + \frac{1}{2}\Delta_z\sigma_z + \frac{\alpha}{\hbar}\left(\Pi_x\sigma_y - \Pi_y\sigma_x\right), \quad (S1)$$

where $\vec{\Pi} = \vec{p} + e\vec{A}$, $\vec{A} = \left(-B_y, B_x, 0\right)/2$, $m^*$ is the electron effective mass, $\Delta_z = g\mu_B B$ is the Zeeman energy, g is the Landé g-factor, and $\alpha$ is the Rashba SOC strength. We express the Hamiltonian in terms of dimensionless operators,

$$\hat{a} = \frac{1}{\sqrt{2}}\left[\left(\frac{x}{2} + \frac{\partial}{\partial x}\right) - i\left(\frac{y}{2} + \frac{\partial}{\partial y}\right)\right], \quad \hat{a}^\dagger = \frac{1}{\sqrt{2}}\left[\left(\frac{x}{2} + \frac{\partial}{\partial x}\right) + i\left(\frac{y}{2} + \frac{\partial}{\partial y}\right)\right],$$

satisfying the *bosonic* commutator $[\hat{a}, \hat{a}^\dagger] = 1$. Then,

$$H = \hbar\omega_c\left(\hat{a}^\dagger\hat{a} + \frac{1}{2}\right) + \frac{1}{2}\Delta_z\sigma_z - \frac{\alpha}{\sqrt{2}l_B}\left(\hat{a}\sigma_+ + \hat{a}^\dagger\sigma_-\right), \quad (S2)$$

where $\sigma_+ = \sigma_x + i\sigma_y$ and $\sigma_- = \sigma_x - i\sigma_y$ are spin ladder operators obeying commutation relations $\{\sigma_+, \sigma_-\} \propto 1$ and $\{\sigma_\pm, \sigma_\pm\} = 0$ that are similar to those of fermionic creation and annihilation operators. We diagonalize the Hamiltonian, using its constant of motion $\hat{N} = \hat{a}^\dagger \hat{a} + \frac{1}{2}\sigma_z$. An eigenstate of $\hat{N}$ is a linear combination of $\{|n+1, -\rangle, |n, +\rangle\}$, where $|n, \pm\rangle$ is an eigenstate of $\hat{N}$ with Landau-level index $n$ and spin-$z$ component $\pm$. In the subspace of a given $n \geq 0$, the Hamiltonian is represented as a $2 \times 2$ matrix,

$$H = (n+1)\hbar\omega_c \mathbb{I} - \frac{2\sqrt{n+1}\alpha}{\sqrt{2}l_B}\sigma_x + \frac{1}{2}(\hbar\omega_c - \Delta_z)\sigma_z,$$

which can be easily diagonalized. We find the eigenenergy of $|n = 0,1,2, \ldots, \sigma = \pm\rangle$ as

$$E_{(n,\sigma)} = \left(n + \frac{\sigma}{2} + \frac{1}{2}\right)\hbar\omega_c - \sigma\sqrt{\left(n + \frac{\sigma}{2} + \frac{1}{2}\right)\frac{2\alpha^2}{l_B^2} + \frac{1}{4}(\hbar\omega_c - \Delta_z)^2}. \quad \text{(S3)}$$

Next, we derive Eq. (1), the eigenenergy of Hamiltonian (S2) in the Paschen-Back (PB) regime of a sufficiently strong magnetic field B, treating the last term of SOC with $\alpha$ as a perturbation of the other terms of (S2). Then, the Hamiltonian is decomposed into $H = H_0 + V_{SO}$, the bare part and the perturbation part. The bare Hamiltonian is diagonalized as $H_0|n, \sigma\rangle = E_{(n,\sigma)}^{(0)}|n, \sigma\rangle$, where $E_{(n,\sigma)}^{(0)} \equiv \left(n + \frac{1}{2}\right)\hbar\omega_c + \sigma\frac{\Delta_z}{2}$. The first order energy correction by the perturbation $V_{SO} \equiv \frac{\alpha}{\sqrt{2}l_B}(\hat{a}\sigma_+ + \hat{a}^\dagger \sigma_-)$ vanishes, $E_{(n,\sigma)}^{(1)} = \langle n, \sigma|V_{SO}|n, \sigma\rangle = 0$, differently from the PB effects in atoms [6]. So the leading-order perturbation correction is of the second order,

$$E_{(n,\sigma)}^{(2)} = \sum_{m \neq n, \sigma' = \pm} \left\langle n, \sigma \left| V_{SO}\left(\frac{|m, \sigma'\rangle\langle m, \sigma'|}{E_{(n,\sigma)}^{(0)} - E_{(m,\sigma')}^{(0)}}\right) V_{SO} \right| n, \sigma \right\rangle = -2\sigma\frac{2n + \sigma + 1}{1 - \frac{\Delta_z}{\hbar\omega_c}}\Delta_{SO} = -\sigma\Delta_{PB},$$

where $\Delta_{SO} = \frac{m^*\alpha^2}{2\hbar^2}$. We then find the energy spectrum of the PB regime in Eq. (1) of the main text,

$$E_{(n,\sigma)} \approx \left(n + \frac{1}{2}\right)\hbar\omega_c + \frac{\sigma}{2}\Delta_z - \sigma\Delta_{PB}. \quad \text{(S4)}$$

This result is also obtained by Taylor expanding Eq. (2) as a power series of $\alpha$ at $\alpha = 0$. The perturbation result in Eq. (S4) (namely Eq. (1) of the main text) is valid for our nanowires at B >

2.46 T and with small Landau level index of $n = 0, 1$. At $B > 2.6$ T, the percentage of deviation of the perturbation result in Eq. (S4) from the exact value from Eq. (S3) is less than 6%, which is obtained with $\alpha = 3.6 \times 10^{-11}\text{eV} \cdot \text{m}$, $g = 20$, and $m^* = 0.025 m_e$.

We have two remarks. First, the last term of Eq. (S4), the perturbation correction, is interpreted as the effective Rashba SOC dressed by the Landau orbital motion and the Zeeman spin splitting. Second, the ground state $|n = 0, -\rangle$ is special in the perturbation expansion. There is no correction to the ground-state energy for all perturbation orders, since the ground state $|n = 0, -\rangle$ satisfies $\frac{\alpha}{\sqrt{2}l_{\text{B}}}(\hat{a}\sigma_+ + \hat{a}^\dagger\sigma_-)|0, -\rangle = 0$. Namely, there is no contribution from the Rashba SOC to the ground state $|n = 0, -\rangle$ of the bare Hamiltonian describing the Landau orbital motion and the Zeeman splitting, or the effects of the Landau orbital motion and the Zeeman spin splitting on the dressed Rashba SOC cancel each other

## 3. Numerical simulation by the recursive Green function method

In this subsection, we consider the modification of the energy spectrum in Eq. (S2) by effects of the finite lateral width W. To take the effects into account, we consider a tight-binding model, discretizing the continuum Hamiltonian in Eq. (S1)

$$H_{\text{lattice}} = \sum_{m=1}^{N_y} \sum_{n=-\infty}^{\infty} \left( C_{n,m}^\dagger T_0 C_{n,m} + C_{n+1,m}^\dagger T_x C_{n,m} + C_{n,m+1}^\dagger T_y C_{n,m} + \text{h.c.} \right),$$

where $C_{n,m}^\dagger = (C_{n,m,\uparrow}^\dagger C_{n,m,\downarrow}^\dagger)$, $T_0 = 2t_0\mathbb{I} + \Delta_z\hat{n} \cdot \vec{\sigma}$, $T_x = -t_0 e^{i\varphi}\mathbb{I} + it_1 e^{i\varphi}\sigma_y$, $T_y = -t_0\mathbb{I} - it_1\sigma_x$, $t_0 = \hbar^2/(2m^*a^2)$, $t_1 = \alpha/(2a)$, $\varphi = eBa^2/\hbar$, and $a = 0.5 \times 10^{-9}$ m. We choose $N_y = 88$ which corresponds to $W = 44$ nm. Based on this Hamiltonian and utilizing the recursive Green function method [7], we obtain the conductance map in Figure 4. .

We present the numerical result about the wave function of the Hamiltonian for the lattice of finite width W. The result is drawn in Figure S1. As shown in the figure (see the region of small k), quantized Landau levels are formed in the bulk of the lattice. It is easy to see that the result is different from the case without the magnetic field.

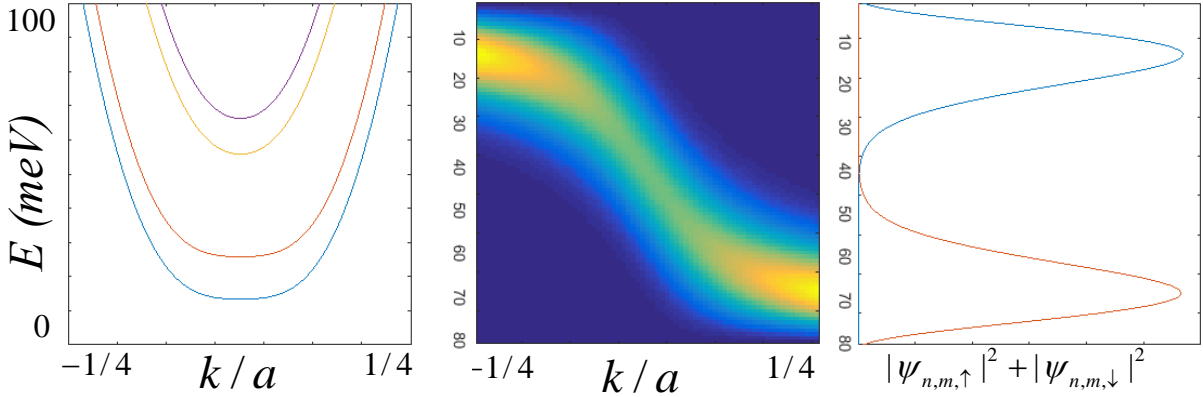

Figure S1. Left panel: Landau levels at $B = 9$ T as a function of longitudinal wave vector $k$. Middle: Color plot of $\left|\psi_{n,m,\uparrow}\right|^2 + \left|\psi_{n,m,\downarrow}\right|^2$ as a function of the longitudinal wave vector $k$ (x axis) and the lattice site index $m$ (i.e. the coordinate) in the lateral direction, where $\psi_{n,m,\sigma}$ is the normalized wave function of the state with longitudinal wave vector k and spin $\sigma$ in the lowest Landau level at lattice site $(n, m)$ and n is the site index in the longitudinal direction. Right: $\left|\psi_{n,m,\uparrow}\right|^2 + \left|\psi_{n,m,\downarrow}\right|^2$ at $k = 1/4$ (see the red line) and $k = -1/4$ (blue) as a function of $m$. The result shows that the wave function is well localized within the magnetic length.

## Section 2. Growth of InAs nanowires

Catalyst-free InAs nanowires are grown on a p-type Si (111) wafer in a horizontal reactor MOCVD system (AIXTRON Inc.). At a proper growth condition, InAs forms isolated islands on Si surface because of large lattice mismatch strain (11%) with Si. In this condition, continued crystallization of the InAs proceeds primarily on <111> surface because of the lowest surface energy in that direction, thus forming one-dimensional crystal structure. Contrary to conventional VLS method, there is no any disorder effect arising from the diffusion of the catalyst metal atoms into the nanowires. The nanowires exhibit a zinc-blende crystal structure but a large number of stacking faults was observed along the nanowire growth direction (i.e. <111>) as shown in the inset of Figure S2. Nevertheless, no misfit dislocation was found across the InAs nanowire. The nanowires have nearly uniform heights of ~20 μm with diameter of ~90 nm with well-controlled growth conditions such as temperature and growth time (see Figure S2).

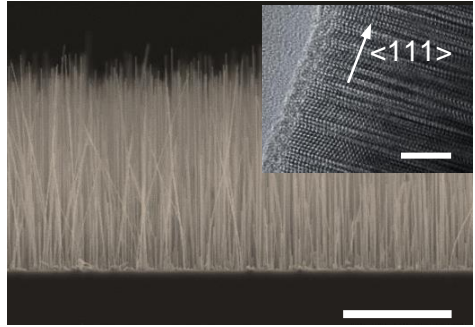

Figure S2. Scanning electron microscope (SEM) image of grown InAs nanowires on a Si(111) wafer, which shows nearly uniform heights of ~20 μm with diameter of ~90 nm. Scale bar is 10 μm. Inset: a high-resolution transmission electron microscope image of a grown InAs nanowire. Scale bar is 5 nm.

## Section 3. Comsol simulation

To simulate the electric potential profile of the device, we performed three-dimensional simulations based on finite element analysis using COMSOL Multiphysics 4.2 software (see Figure S3). Here, electrostatics module was used in the consideration of electric conductivity and relative permittivity of materials. Figure S3(a) shows the simulated device geometry, where InAs nanowire is assumed to be a hexagonal cylinder with thickness of 70 nm. The distance between the source and drain contacts is 85 nm. In this calculation, to reduce the computation time, we assumed the thickness of the dielectric $SiO_2$ (metal electrodes) is 50 (80) nm, which is smaller than the actual value 300 (120) nm. Figure S3(b) illustrates the simulated electric potential profile. Figure S3(c) shows a slice cut of potential profile along the channel at the top-center ($z = 120$ nm, $x = 75$ nm) of nanowire. It shows an arch shape with respect to the middle of the channel, where a negative electrical potential is maximum. Under a relevant gate voltage condition, this leads to a sufficiently thin channel to form an effective two-dimensional electron gas system near the middle of the channel.

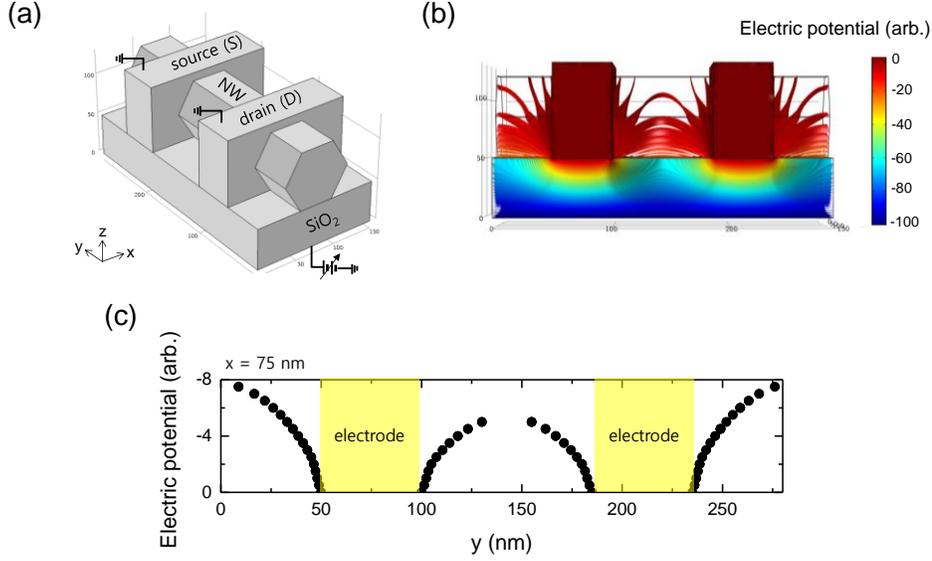

Figure S3. (a) Outline of the device for the three-dimensional finite-element method calculations of InAs nanowire field-effect transistor. The length (diameter) of nanowire, thicknesses of metal for source and drain and thickness of SiO₂ are 85 (70) nm, 80 nm and 50 nm, respectively. The source and drain are grounded. The bottom of SiO₂ layer is electrically biased as a back-gate field. (b) Simulated electric potential profile. (c) Slice cut of potential profile along the channel at the top-center (*z* = 120 nm, *x* = 75 nm) of nanowire.

## Section 4. Fabry-Perot interference

With nearly transparent contacts, InAs nanowire devices have shown the Fabry-Perot interference effect [8]. A constructive interference can occur when an accumulated phase of electrons reflected by two contacts in a nanowire satisfies a condition of $\Delta k \times 2L = 2\pi n$, where, $\Delta k$ is a deviation of a wave number, $L$ a channel length, and $n$ an integer number (see Figure S4). To get the energy scale of the Fabry-Perot interference, we consider a simple parabolic band structure represented by $E = \hbar^2 k^2 / 2m^*$, as shown in Figure S4. Here, $m^*$ is the effective mass of electron in InAs nanowires. To calculate the energy variation for $\Delta k$, we set a following equation: $\int_{E_F}^{\mu} dE = \frac{\hbar^2}{m^*} \int_{k_F}^{k_F + \Delta k} k \, dk$. Then, we get $E_{FP} \sim \frac{h v_f}{2L}$ with $\mu = E_{FP}$ and $\Delta k = \frac{\pi}{L}$, where $v_f$ is the Fermi velocity and $E_{FP}$ is the energy scale corresponding to the Fabry-Perot interference.

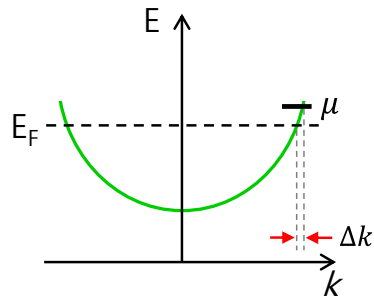

Figure S4. Parabolic band diagram for calculation of the energy scale of the Fabry-Perot interference.

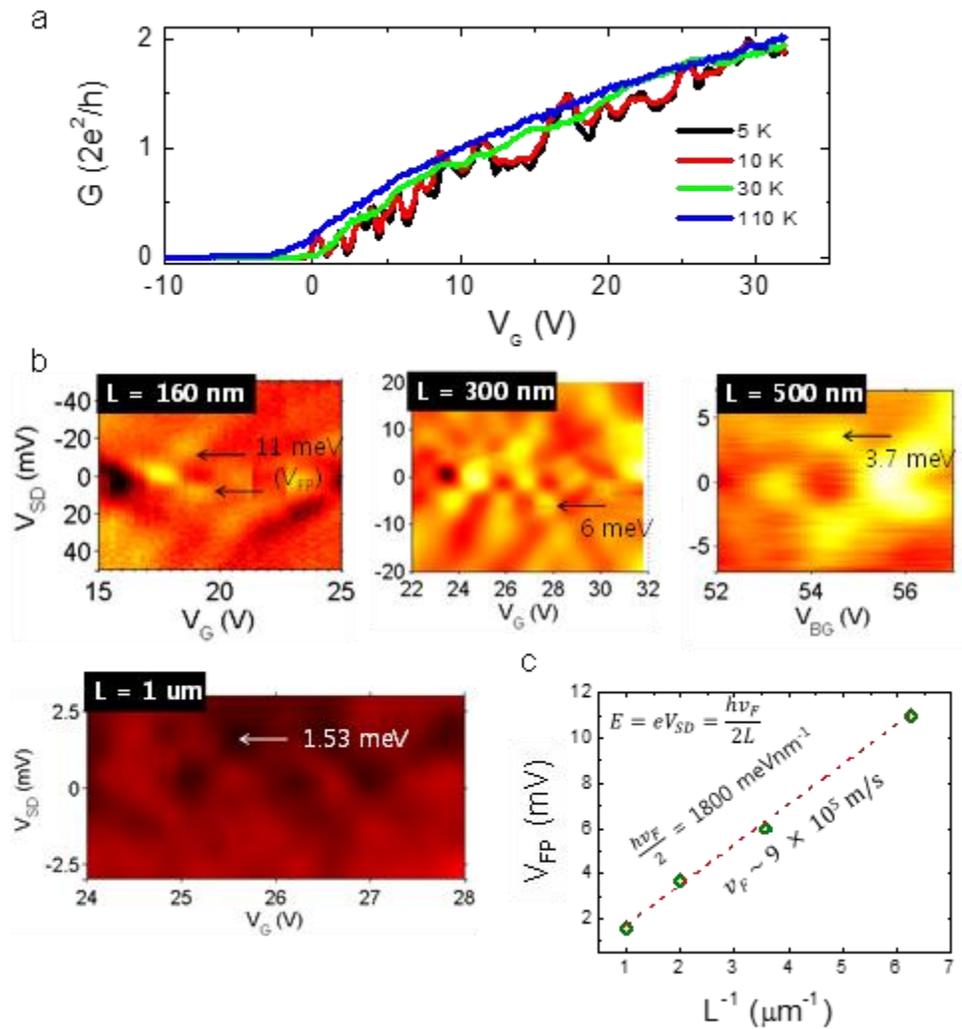

Figure S5. (a) Conductance as a function of gate voltage ($G$-$V_G$) for various temperatures of NW2 ($L$ = 160 nm). At $T$ = 5 K, the curve shows a quasi-periodic oscillation behavior, which is originated by the Fabry-Perot interference. With increasing temperature, the interference effect is smeared out. (b) d$G$/d$V_g$

maps as a function of $V_{SD}$ and $V_G$ for $L$ = 160 nm, 300 nm, 500 nm and 1 μm, where bright and dark regions indicate relatively high and low d$G$/d$V_g$ values, respectively. All channels show a checkerboard pattern representing the Fabry-Perot interference. The energy scale of the Fabry-Perot ($V_{FP}$) is depicted by arrows in each panel. (c) $V_{FP}$ as a function of $L^{-1}$, where the slope of the scattered points provides a value of the Fermi velocity, $v_F$ of InAs nanowire as ~9×10$^5$ m/s. This value is consistent with a previous report [9].

## Section 5. Comparison between experiments and theory for with and without dressed SOC effect at negative $V_{sd}$.

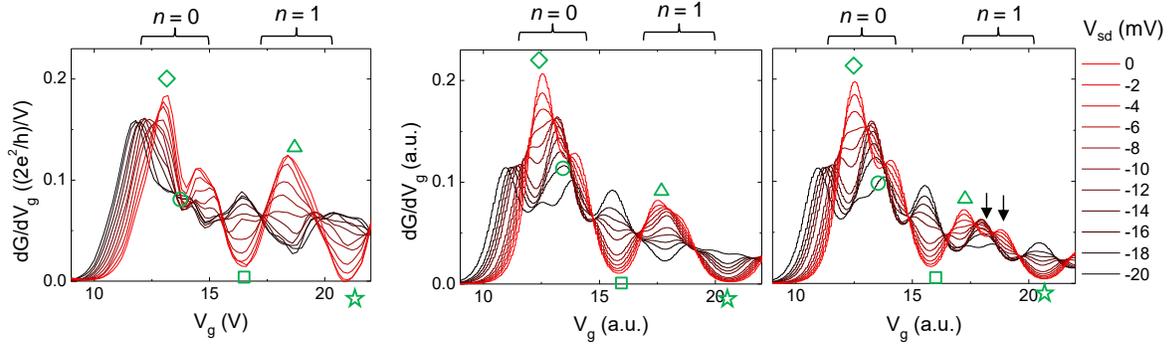

Figure S6. (a) From Left to Right: The same plots of Figures 5(a)-5(c) in the main text for negative $V_{sd}$ values, respectively, but without data shift to compare the d$G$/d$V_g$ peak amplitudes in the $n$ = 0 and 1 LL regions. The left and middle panels show the same trend in modulations of the dips and peaks as varying $V_{sd}$ in the $n$ =1 LL region, contrary to those in the right panel.